\documentclass[12pt]{article}
\usepackage[centertags]{amsmath}
\usepackage{enumerate}
\usepackage{stmaryrd}
\usepackage{dsfont}
\usepackage{url}
\usepackage{geometry}
\usepackage{hyperref}
\usepackage[latin1]{inputenc}
\usepackage{amsfonts,amssymb,amsmath,amscd,slashed,amsthm}
\usepackage{mathrsfs}
\usepackage{enumerate}
\usepackage[dvips]{graphicx}
\usepackage{color}
\usepackage{verbatim}
\usepackage{psfrag}
\usepackage{epsfig}

\usepackage{epsfig}
\usepackage{epstopdf}

\usepackage{hyphenat}

\usepackage[english]{babel}

\newtheorem{Definition}{Definition}
\newtheorem{Conjecture}{Conjecture}
\newtheorem{Example}{Example}
\newtheorem{Theorem}{Theorem}
\newtheorem{Claim}{Claim}
\newtheorem{Lemma}{Lemma}
\newtheorem{Remark}{Remark}
\newtheorem{Corollary}{Corollary}
\newtheorem{Proposition}{Proposition}

\newenvironment{Proof}{\textbf{Proof}}{ \qed \\}

\newcommand{\sN}{{\mathbb N}}

\newcommand{\sR}{{\mathbb R}}

\newcommand{\B}{\mathcal{B}}

\newcommand{\F}{\mathcal{F}}
\newcommand{\M}{\mathcal{M}}
\newcommand{\Pf}{\mathscr{P}}

\newcommand{\T}{\mathcal{T}}

\newcommand{\X}{\mathcal{X}}
\newcommand{\Y}{\mathcal{Y}}

\begin{document}

\title{On the causality and $K$-causality between measures}

\author{Tomasz Miller\thanks{E-mail: T.Miller@mini.pw.edu.pl} \\
\footnotesize{Faculty of Mathematics and Information Science, Warsaw University of Technology,} \\ \footnotesize{ul. Koszykowa 75, 00-662 Warsaw, Poland,} \\
\footnotesize{Copernicus Center for Interdisciplinary Studies, ul. S{\l}awkowska 17, 31-016 Krak\'ow, Poland}}

\maketitle


\begin{abstract}
Drawing from our earlier works on the notion of causality for nonlocal phenomena, we propose and study the extension of the Sorkin--Woolgar relation $K^+$ onto the space of Borel probability measures on a given spacetime. We show that it retains its fundamental properties of transitivity and closedness. Furthermore, we list and prove several characterizations of this relation, including the `nonlocal' analogue of the characterization of $K^+$ in terms of time functions. This generalizes and casts new light on our earlier results concerning the causal precedence relation $J^+$ between measures.
\end{abstract}

\section{Introduction}

One of the vital fibers of mathematical relativity is the study of the (global) causal properties of spacetimes --- the causality theory. The fundamental role in this theory is played by certain binary relations on a given spacetime, designed to describe which events (e.g. points of spacetime) are causally connected, and which can never constitute links of a cause-effect chain. The two most important such relations --- called $I^+$ and $J^+$ --- involve pairs of events $(p,q)$ which can be connected by means of a chronological curve (what is denoted $p \ll q$) or a causal curve (what is usually denoted $p \preceq q$), respectively. For a concise review of the causality theory, the reader is referred to \cite{MS08} (see also \cite{Beem,HELargeScaleStructure,BN83,Penrose1972,Wald} for the full exposition).

Motivated by the Lorentzian version of noncommutative geometry \cite{PROC2015,F4,CQG2013,SIGMA2014,JGP2015}, where spacetime usually can no longer be regarded as composed of pointlike events, we have recently proposed and studied a natural extension of the relation $\preceq$ onto the space $\Pf(\M)$ of Borel probability measures on a given spacetime $\M$ \cite{EcksteinMiller2015}, which relies on the notion of a \emph{coupling} drawn from the optimal transport theory \cite{UsersGuide,Villani2008}. Given a pair of measures $\mu,\nu \in \Pf(\M)$, we call $\omega \in \Pf(\M^2)$ a \emph{coupling} of $\mu$ and $\nu$ iff the latter two measures are $\omega$'s marginals, that is iff $\pi^1_\sharp \omega = \mu$ and $\pi^2_\sharp \omega = \nu$, where $\pi^i: \M^2 \rightarrow \M$, $i=1,2$ denote the canonical projections. With $\Pi(\mu,\nu)$ denoting the set of all such couplings, we have put forward the following definition:
\begin{Definition}
\label{causality_def}
Let $\M$ be a~spacetime. For any $\mu,\nu \in \Pf(\M)$ we say that $\mu$ \emph{causally precedes} $\nu$ (symbolically $\mu \preceq \nu$) iff there exists a \emph{causal coupling} of $\mu$ and $\nu$, by which we mean any $\omega \in \Pi(\mu,\nu)$ such that $\omega(J^+) = 1$.
\end{Definition}
Let us emphasize that the quantity $\omega(J^+)$ is well-defined, because $J^+$ is a Borel subset of $\M^2$ \cite[Section 3]{EcksteinMiller2015}. The name ``causal coupling'' was independently coined in \cite{Suhr2016}.

This definition rigorously encapsulates the following common intuition: Infinitesimal portions of probability distributions can propagate in spacetime only along future-directed causal curves.
The thus obtained formalism can be successfully used to model the causal time-evolution of spatially distributed physical quantities \cite{Miller2016} and wave packets \cite{2NEW2016}, as well as to quantify the breakdown of causality in various physical models. Although the idea of studying causality for spread objects is by no means new --- see e.g. \cite{Gerlach1969,Gerlach1968,Gromes1970,Hegerfeldt1,Hegerfeldt1985,HegerfeldtFermi,Hegerfeldt2001,Hegerfeldt2} --- the tools of the modern optimal transport theory adapted to the Lorentzian setting allow us to cast some new light on the previous approaches and significantly extend them. See \cite{2NEW2016} for concrete physical applications and a more detailed discussion.

Even though $I^+, J^+$ play the central role in mathematical relativity, these relations are not the only ones studied in causality theory, with the so-called Seifert's relation $J^+_S \supseteq J^+$ providing an important example \cite{Seifert1971}. Unfortunately, both $J^+$ and $J^+_S$ exhibit certain unfavourable features. Namely, the relation $J^+$, although always transitive, need not be closed topologically, whereas $J^+_S$, albeit closed, need not be transitive. One might thus look for the smallest closed and transitive relation containing $J^+$ --- the project first accomplished by Sorkin and Woolgar \cite{SorkinWoolgar} who called such relation $K^+$. It turns out that $K^+$ is additionally antisymmetric (and hence a partial order) precisely when the spacetime is stably causal, in which case $K^+ = J^+_S = \overline{J^+}$ \cite{Minguzzi2009}. Furthermore, $K^+$ possesses the following characterization in terms of \emph{time functions} \cite{MinguzziUtilities}.
\begin{Theorem}[Minguzzi]
\label{MinguzziThm}
Let $\M$ be a stably causal spacetime and $p, q \in \M$. Then
\begin{align}
\label{MinguzziT}
(p, q) \in K^+ \quad \Leftrightarrow \quad \textnormal{For every time function } \T \quad \T(p) \leq \T(q).
\end{align}
\end{Theorem}

In his work \cite{Minguzzi2009}, E. Minguzzi calls $K^+$ ``one of the most important [relations] for the development of causality theory''. It is thus natural to ask whether the formalism developed in \cite{EcksteinMiller2015} for $J^+$ can be adapted to $K^+$. Indeed, Definition \ref{causality_def} can be immediately modified to extend the relation $K^+$ (instead of $J^+$) onto $\Pf(\M)$ (see Definition \ref{Kcausality_def} below). The aim of this paper is to study thus obtained relation between measures. More concretely, in Section \ref{sec::results} we show that the Sorkin--Woolgar relation, when extended onto $\Pf(\M)$, retains its fundamental properties of transitivity and closedness (Proposition \ref{Prop1}). Then we establish its various characterizations (Theorem \ref{main}), in particular providing the `nonlocal' counterpart of equivalence (\ref{MinguzziT}). All that generalizes and sheds new light on some of the results presented in \cite{EcksteinMiller2015}, what is briefly discussed in Section \ref{sec::conclusions}.

\section{Results}
\label{sec::results}

From now on, the term ``measure'' will always stand for ``Borel probability measure''.

Analogously as for the relations $I^+$ and $J^+$, for any $\X \subseteq \M$ one introduces the notation $K^+(\X) := \pi^2\left((\X \times \M) \cap K^+\right)$ and $K^-(\X) := \pi^1\left((\M \times \X) \cap K^+\right)$. By definition, $K^+$ is closed, and hence for any compact $C \subseteq \M$ the sets $K^\pm(C)$ are closed. More generally, however, for $\X \subseteq \M$ which is only Borel, the sets $K^\pm(\X)$ need not be Borel. Nevertheless, being projections of Borel sets, they are \emph{universally measurable}, which means that for any measure $\mu \in \Pf(\M)$ the sets $K^\pm(\X)$ are Borel up to a $\mu$-negligible set, and therefore the quantity $\mu(K^\pm(\X))$ is well defined \cite{Aliprantis}.

As announced in the Introduction, let us put forward the following definition of the Sorkin--Woolgar relation between measures.
\begin{Definition}
\label{Kcausality_def}
Let $\M$ be a~spacetime. For any $\mu,\nu \in \Pf(\M)$ we say that $\mu$ $K$-\emph{causally precedes} $\nu$ (symbolically $\mu \preceq_K \nu$) iff there exists a $K$-\emph{causal coupling} of $\mu$ and $\nu$, by which we mean any $\omega \in \Pi(\mu,\nu)$ such that $\omega(K^+) = 1$.
\end{Definition}

As a first result, let us observe that the defining properties of the Sorkin--Woolgar relation --- closedness and transitivity --- still hold after extending it onto $\Pf(\M)$.
\begin{Proposition}
\label{Prop1}
The relation $\preceq_K$ on $\Pf(\M)$ is reflexive and transitive, as well as closed, by which we mean that the set $\{ (\mu,\nu) \in \Pf(\M)^2 \, | \, \mu \preceq_K \nu \}$ is closed in $\Pf(\M)^2$ endowed with the (product) narrow topology.
\end{Proposition}
\begin{Proof}\textbf{.}
Reflexivity and transitivity can be shown exactly as in the proof of \cite[Theorem 11]{EcksteinMiller2015} with $\preceq$ and $J^+$ replaced by $\preceq_K$ and $K^+$.

As for the closedness, take any sequences $(\mu_n), (\nu_n) \subseteq \Pf(\M)$ satisfying $\mu_n \preceq_K \nu_n$ for all $n \in \sN$, and suppose that they converge, respectively, to some $\mu, \nu \in \Pf(\M)$ in the narrow topology, i.e. that
\begin{align*}
    \forall \, f \in C_b(\M) \quad \int\limits_\M f d\mu_n \rightarrow \int\limits_\M f d\mu \quad \textrm{and} \quad \int\limits_\M f d\nu_n \rightarrow \int\limits_\M f d\nu.
\end{align*}
What we need to prove is that $\mu \preceq_K \nu$. To this end, let $\omega_n$ be the $K$-causal coupling of $\mu_n$ and $\nu_n$ for every $n \in \sN$. We claim that the sequence $(\omega_n) \subseteq \Pf(\M^2)$ has a narrowly convergent subsequence. Indeed, the sets $\{\mu_n\}_{n \in \sN}$, $\{\nu_n\}_{n \in \sN}$ are relatively narrowly compact and hence tight by the Prokhorov theorem. But this implies that the set $\bigcup\limits_{i,j \in \sN} \Pi(\mu_i,\nu_j)$ is also tight \cite[Lemma 4.4]{Villani2008} and hence relatively compact. Since $(\omega_n)$ is contained in $\bigcup\limits_{i,j \in \sN} \Pi(\mu_i,\nu_j)$, it possesses a subsequence (which we keep denoting as $(\omega_n)$) convergent to some $\omega \in \Pf(\M^2)$.

It now remains to show that $\omega$ is a $K$-causal coupling of $\mu$ and $\nu$. In order to prove that $\omega \in \Pi(\mu,\nu)$, observe first that for any $g \in C_b(\M)$
\begin{align*}
\int\limits_\M g \, d\left(\pi^1_\sharp \omega\right) = \int\limits_\M (g \circ \pi^1) d\omega = \lim\limits_{n \rightarrow +\infty} \int\limits_\M (g \circ \pi^1) d\omega_n = \lim\limits_{n \rightarrow +\infty} \int\limits_\M g \, d\mu_n = \int\limits_\M g \, d\mu,
\end{align*}
which proves that $\pi^1_\sharp \omega = \mu$. The equality $\pi^2_\sharp \omega = \nu$ is proven analogously. Finally, in order to show that $\omega(K^+) = 1$, invoke the portmanteau theorem, on the strength of which $\omega(F) \geq \limsup\limits_{n \rightarrow +\infty} \omega_n(F)$ for any closed $F \subseteq \M^2$, and simply take $F := K^+$.
\end{Proof}

We now provide several characterizations of the relation $\preceq_K$ for measures, what constitutes the main result of the paper.
\begin{Theorem}
\label{main}
Let $\M$ be a spacetime. For any $\mu,\nu \in \Pf(\M)$ consider the following list of conditions.
\begin{enumerate}[1{$^\bullet$}]
\item $\mu \preceq_K \nu$.
\item For any compact subset $\ C \subseteq \M$
\begin{align}
\label{main1}
\mu(K^+(C)) \leq \nu(K^+(C)).
\end{align}
\item For any Borel subset $\X \subseteq \M$ such that $K^+(\X) \subseteq \X$
\begin{align}
\label{main2}
\mu(\X) \leq \nu(\X).
\end{align}
\item For any time function $\T$ and any $\alpha \in \sR$
\begin{align}
\label{main3}
\mu\left(\T^{-1}((\alpha, +\infty))\right) \leq \nu\left(\T^{-1}((\alpha, +\infty))\right).
\end{align}
\item For any bounded time function $\T$
\begin{align}
\label{main4}
\int_\M \T d\mu \leq \int_\M \T d\nu.
\end{align}
\end{enumerate}
Then we have the following equivalences and implications $1^\bullet \ \Leftrightarrow \ 2^\bullet \ \Leftrightarrow \ 3^\bullet \ \Rightarrow \ 4^\bullet \ \Rightarrow \ 5^\bullet$. Moreover, if $\M$ is causally continuous, then all above conditions are equivalent.
\end{Theorem}

Observe that conditions $4^\bullet$ and $5^\bullet$ implicitly assume that $\M$ is stably causal. Condition $5^\bullet$ is a natural extension of Minguzzi's condition (\ref{MinguzziT}) onto measures. Notice, however, that we assume here the causal continuity of $\M$ for the equivalence $1^\bullet \Leftrightarrow 5^\bullet$ to hold. 

All above conditions (with the exception of $4^\bullet$) are analogues of those studied in \cite{EcksteinMiller2015} in the context of the relation $J^+$, and in fact they reduce to them for the case of causally simple spacetimes. However, the properties of $K^+$ allow these analogues to work in a broader class of spacetimes. Some parts of the following proof simply mimic the argumentation presented in \cite{EcksteinMiller2015}, while others are significantly different, thus offering an alternative way to reach the results of the cited work. In particular, the crucial role will be played by the following fact, adapted from a more general result obtained by Suhr, cf. \cite[Theorem 2.5]{Suhr2016}
\begin{Theorem}[Suhr]
\label{Suhrowe}
For any spacetime $\M$ and any $\mu, \nu \in \Pf(\M)$ the following conditions are equivalent
\begin{enumerate}[i)]
\item $\mu \preceq_K \nu$.
\item For any Borel $\B \subseteq \M$
\begin{align*}
\mu(\B) \leq \nu(K^+(\B)) \quad \textnormal{and} \quad \mu(K^-(\B)) \geq \nu(\B).
\end{align*}
\end{enumerate}
\end{Theorem}

We will also need the following lemma, the proof of which is a straightforward adaptation of that of \cite[Proposition 1]{EcksteinMiller2015}.
\begin{Lemma}
\label{lemacik}
For any spacetime $\M$ and any $\X \subseteq \M$ denote $\X^c := \M \setminus \X$. We have the equivalence of inclusions $K^+(\X) \subseteq \X \ \Leftrightarrow \ K^-(\X^c) \subseteq \X^c$.
\end{Lemma}

\begin{Proof} \textbf{of Theorem \ref{main}.}
$1^\bullet \ \Rightarrow \ 2^\bullet$ By the closedness of $K^+$, the set $K^+(C)$ is closed, and hence Borel, for any compact $C$. Denoting the characteristic function of $K^+(C)$ by $\chi$, the inequality $\chi(p) \leq \chi(q)$ holds for all $(p,q) \in K^+$ by the transitivity of $K^+$. By assumption, there exists a $K$-causal coupling $\omega \in \Pi(\mu,\nu)$ and one has that
\begin{align*}
\mu(K^+(C)) & = \int\limits_\M \chi(p) d\mu(p) = \int\limits_{\M^2} \chi(p) d\omega(p,q) = \int\limits_{K^+} \chi(p) d\omega(p,q) \leq \int\limits_{K^+} \chi(q) d\omega(p,q)
\\
& = \int\limits_{\M^2} \chi(q) d\omega(p,q) = \int\limits_\M \chi(q) d\nu(q) = \nu(K^+(C)).
\end{align*}

$2^\bullet \ \Rightarrow \ 3^\bullet$ Let the set $\X$ be as specified in $3^\bullet$ and let $C$ be any compact subset of $\X$. Then also $K^+(C) \subseteq K^+(\X) \subseteq \X$ and one has that
\begin{align*}
\mu(C) \leq \mu(K^+(C)) \leq \nu(K^+(C)) \leq \nu(\X),
\end{align*}
where the inequality holds by $2^\bullet$. Taking now the supremum over all compacts $C \subseteq \X$, by the well-known fact that every Borel probability measure on a Polish space is tight \cite{Srivastava} we obtain that
\begin{align*}
\mu(\X) = \sup \{ \mu(C) \, | \, C \subseteq \X \textnormal{ compact} \} \leq \nu(\X).
\end{align*}

$3^\bullet \ \Rightarrow \ 1^\bullet$ It suffices to prove that condition ii) in Theorem \ref{Suhrowe} is satisfied.

To this end, observe that the first inequality in ii) follows directly from $3^\bullet$ with $\X := K^+(\B)$ (up to a $\mu$-neglectable set) and the obvious inequality $\mu(\B) \leq \mu(K^+(\B))$. In order to obtain the second inequality, notice that $3^\bullet$ is in fact equivalent to its ``$K^-$'' counterpart, namely
\begin{enumerate}[1{$^{\prime \bullet}$}]
\setcounter{enumi}{2}
\item For any Borel subset $\Y \subseteq \M$ such that $K^-(\Y) \subseteq \Y$
\begin{align}
\label{main2a}
\mu(\Y) \geq \nu(\Y).
\end{align}
\end{enumerate}
Indeed, to move between conditions $3^\bullet$ and $3^{\prime \bullet}$, simply take $\Y = \X^c$ and invoke Lemma \ref{lemacik}.

The second inequality in condition ii) of Theorem \ref{Suhrowe} follows from $3^{\prime \bullet}$ with $\Y := K^-(\B)$ (up to a $\nu$-neglectable set) and the obvious inequality $\nu(K^-(\B)) \geq \nu(\B)$.
\\

$3^\bullet \ \Rightarrow \ 4^\bullet$ Take any time function $\T$ (from now on, we implicitly assume that $\M$ is stably causal) and any $\alpha \in \sR$. On the strength of $3^\bullet$, we only have to show that $K^+\left(\T^{-1}((\alpha, +\infty))\right) \subseteq \T^{-1}\left((\alpha, +\infty)\right)$.

To this end, let $q \in \M$ be such that there exists $p \in \T^{-1}\left((\alpha, +\infty)\right)$ with $(p,q) \in K^+$. By Theorem \ref{MinguzziThm}, we immediately obtain that $\T(q) \geq \T(p) > \alpha$ and so $q \in \T^{-1}\left((\alpha, +\infty)\right)$ as well.
\\

$4^\bullet \ \Rightarrow \ 5^\bullet$ Fix a bounded time function $\T$. Reasoning similarly as in the proof of \cite[Theorem 7, $3^\bullet \Rightarrow 2^\bullet$]{EcksteinMiller2015}, denote $m := \inf \T$ and $M := \sup \T$ and for every $n \in \sN$ define the simple function
\begin{align*}
s_n := m + \sum\limits_{k=1}^{n-1} \tfrac{M-m}{n} \chi_{\F^{(n)}_k},
\end{align*}
where $\F^{(n)}_k := \T^{-1}\left((m + k \tfrac{M-n}{n}, +\infty)\right)$, $k=1,2,\ldots, n-1$. and the functions $\chi_{\F^{(n)}_k}$ denote their respective characteristic functions. On the strength of $4^\bullet$, we have that
\begin{align}
\label{45a}
\int\limits_\M s_n d\mu = m + \sum\limits_{k=1}^{n-1} \tfrac{M-m}{n} \mu(\F^{(n)}_k) \leq m + \sum\limits_{k=1}^{n-1} \tfrac{M-m}{n} \nu(\F^{(n)}_k) = \int\limits_\M s_n d\nu.
\end{align}
The functions $s_n$ are designed so as to satisfy $0 < \T - s_n \leq \tfrac{M-m}{n}$ for all $n \in \sN$. The somewhat technical justification, which we skip here, can be found in the above mentioned proof in \cite{EcksteinMiller2015}. This property allows us to go with $n$ to infinity in (\ref{45a}) by Lebesgue's dominant convergence theorem, what yields (\ref{main4}).
\\

$5^\bullet \ \Rightarrow \ 2^\bullet$ Fix a compact set $C$ and define $\Y := K^+(C)^c$. By the closedness of $K^+(C)$ and Lemma \ref{lemacik}, $\Y$ is an open set with the property that $K^-(\Y) \subseteq \Y$.

We go along the lines of the suitably modified proof of \cite[Theorem 7, $2^\bullet \Rightarrow 3^\bullet$]{EcksteinMiller2015}. To begin with, let $\eta \in \Pf(\M)$ be an \emph{admissible measure}, which by definition satisfies $\eta(U) > 0$ for all open $U \subseteq \M$ and $\eta(\partial I^{\pm}(p)) = 0$ for all $p \in \M$ \cite[Definition 3.19]{Beem}. For any $\lambda \in (0,1]$ the measure $\eta_\lambda := \lambda \eta + (1 - \lambda) \eta(\, . \, \cap \Y)$ is another admissible measure. Its associated future volume function $t^+_\lambda$, defined for any $p \in \M$ via
\begin{align*}
    t^+_\lambda(p) := - \eta_\lambda(I^+(p)) = -\eta(I^+(p) \cap \Y) - \lambda \eta(I^+(p) \setminus \Y)
\end{align*}
is a (bounded) \emph{time} function provided $\M$ is causally continuous (this is where this assumption comes into play). Take now any sequence of strictly increasing functions\footnote{For instance, taking $\varphi_n(x) := - \tfrac{1}{2} + \tfrac{1}{2} \tanh(n^2 x + n)$ would do.} $(\varphi_n) \subseteq C_b(\sR)$ pointwise convergent to the negative characteristic function of the interval $(-\infty,0)$, denoted here as $-\chi_{<0}$. By $5^\bullet$, for all $\lambda \in (0,1]$ and all $n \in \sN$ it is true that
\begin{align*}
    \int\limits_\M \varphi_n\left( t^+_\lambda(p) \right)d\mu(p) \leq \int\limits_\M \varphi_n\left( t^+_\lambda(p) \right)d\nu(p).
\end{align*}
Using the dominant convergence theorem twice, first for taking $\lambda \rightarrow 0^+$, and then $n \rightarrow +\infty$, we obtain that
\begin{align}
\label{45b}
    -\int\limits_\M \chi_{<0}\left( -\eta(I^+(p) \cap \Y) \right) d\mu(p) \leq -\int\limits_\M \chi_{<0}\left( -\eta(I^+(p) \cap \Y) \right) d\nu(p).
\end{align}
It is now crucial to observe that the map $p \mapsto \eta(I^+(p) \cap \Y)$ is \emph{positive} on $\Y$ and \emph{zero} on $\Y^c$. Indeed, for $p \in \Y$ the set $I^+(p) \cap \Y$ is open and nonempty by the openness of $\Y$, and therefore $\eta(I^+(p) \cap \Y) > 0$ by the very definition of an admissible measure. Conversely, if $\eta(I^+(p) \cap \Y) > 0$ then there exists $q \in \Y$ such that $p \ll q$, which in turn means that $p \in I^-(q) \subseteq K^-(q) \subseteq K^-(\Y) \subseteq \Y$.

By the above observation, the integrands in (\ref{45b}) are nothing but the characteristic function of $\Y$ and hence (\ref{45b}) boils down to $-\mu(\Y) \leq -\nu(\Y)$, which in turn yields $2^\bullet$, because
\begin{align*}
\mu(K^+(C)) = \mu(\Y^c) = 1-\mu(\Y) \leq 1-\nu(\Y) = \nu(\Y^c) = \nu(K^+(C)).
\end{align*}
\end{Proof}

\begin{Remark}
\label{rem1}
If one replaces in condition $5^\bullet$ the term ``bounded'' with ``$\mu$- and $\nu$-integrable'', and/or the term ``time'' with ``temporal'', ``smooth time'', ``smooth causal'' or ``continuous causal'', thus obtained condition is equivalent with $5^\bullet$.
\end{Remark}
\begin{Proof}\textbf{.}
To begin with, let $\tau$ stand for any term from the set $\{$temporal, smooth time, time, smooth causal, continuous causal$\}$. Let us assume that inequality (\ref{main4}) holds for all bounded $\tau$ functions and take any $\mu$- and $\nu$-integrable $\tau$ function $\T$. For every $n \in \sN$ consider the bounded function $n \tanh (\tfrac{1}{n}\T)$, which is also $\tau$. By assumption,
\begin{align}
\label{rem2a}
\int\limits_\M n \tanh \left(\tfrac{1}{n}\T(p)\right) d\mu(p) \leq \int\limits_\M n \tanh \left(\tfrac{1}{n}\T(p)\right) d\nu(p).
\end{align}
Notice now that $|n \tanh (\tfrac{1}{n}\T)| \leq \T$ and that $n \tanh (\tfrac{1}{n}\T) \rightarrow \T$ pointwise, and hence, by the dominant convergence theorem, inequality (\ref{rem2a}) becomes (\ref{main4}) as $n$ approaches infinity.

The converse implication holds trivially, because every bounded $\tau$ function is $\mu$-integrable for any $\mu \in \Pf(\M)$.

We now move to proving that replacing ``time'' in $5^\bullet$ with any term from the set $\{$temporal, smooth time, time, smooth causal, continuous causal$\}$ yields a condition equivalent to $5^\bullet$. Notice that the ``continuous causal'' version of $5^\bullet$ \emph{implies} all other versions, whereas the ``temporal'' version of $5^\bullet$ \emph{is implied by} every other version. We thus only have to show that the ``temporal'' version implies the ``continuous causal'' one.

To this end, assume that $f \in C_b(\M)$ is a causal function and let $t$ be a fixed bounded time function. Then for any $n \in \sN$ $f + \tfrac{1}{n} t$ is another bounded time function, which in turn can be approximated by a sequence of temporal functions $(\T_{n,m})_{m \in \sN}$ such that $| f + \tfrac{1}{n} t - \T_{n,m} | < \tfrac{1}{m}$ everywhere (cf. \cite[Corollary 5.4 and the subsequent comments]{CGM2016}). If we assume that the ``temporal'' version of $5^\bullet$ holds, then we can write that for any $n,m \in \sN$
\begin{align}
\label{rem2b}
\int\limits_\M \T_{n,m} \, d\mu \leq \int\limits_\M \T_{n,m} \, d\nu.
\end{align}
With the aid of Lebesgue's dominant convergence theorem, we can go with $m$, and then with $n$ to infinity, thus obtaining $\int_\M f d\mu \leq \int_\M f d\nu$.
\end{Proof}

\begin{Remark}
\label{rem2}
In condition $4^\bullet$ one can equivalently use the \emph{closed} half-lines. In other words, $4^\bullet$ is equivalent to the following condition:
\begin{enumerate}[1{$^{\prime \bullet}$}]
\setcounter{enumi}{3}
\item For any time function $\T$ and any $\alpha \in \sR$
\begin{align}
\label{main3a}
\mu\left(\T^{-1}([\alpha, +\infty))\right) \leq \nu\left(\T^{-1}([\alpha, +\infty))\right).
\end{align}
\end{enumerate}
\end{Remark}
\begin{Proof}\textbf{.}
To move between conditions \ref{main3} and \ref{main3a}, one can simply use the facts that
\begin{align*}
\T^{-1}([\alpha, +\infty)) = \bigcap\limits_{n=1}^\infty \T^{-1}\left((\alpha-\tfrac{1}{n}, +\infty)\right) \quad \textrm{and} \quad \T^{-1}((\alpha, +\infty)) = \bigcup\limits_{n=1}^\infty \T^{-1}\left([\alpha+\tfrac{1}{n}, +\infty)\right)
\end{align*}
and employ the continuity of measures from below and above.
\end{Proof}

\section{Conclusions}
\label{sec::conclusions}

In this work we have defined the natural extension of the Sorkin--Woolgar relation $K^+$ onto $\Pf(\M)$ and studied its basic properties and characterizations. Since $K^+ = J^+$ for causally simple spacetimes, many of the results obtained here reduce to those known from \cite{EcksteinMiller2015}. However, the current paper yields two new properties of the causal precedence relation $\preceq$ on $\Pf(\M)$ for causally simple $\M$. Namely, by Proposition \ref{Prop1} the relation $\preceq$ on $\Pf(\M)$ is closed, whereas by Theorem \ref{main} conditions $4^\bullet$ and $4^{\prime \bullet}$ provide additional characterizations of $\mu \preceq \nu$ in terms of inverse images of half-lines under time functions. Finally, observe that Remark \ref{rem1}, making reference neither to $\preceq$ nor $\preceq_K$, generalizes \cite[Theorem 6]{EcksteinMiller2015}.

There are still some natural questions concerning the extension of $K^+$ onto $\Pf(\M)$ to be addressed. In the future work we shall investigate e.g. the \emph{antisymmetry} of $\preceq_K$ for measures, which for pointlike events is known to hold exactly in stably causal spacetimes \cite{Minguzzi2009}. The question is whether stable causality already guarantees that $\mu \preceq_K \nu \preceq_K \mu \ \Rightarrow \ \mu = \nu$ or rather one has to impose some stronger requirements on the causal properties of $\M$, such as causal continuity.

\vspace{6pt}

\section*{Acknowledgements}
The author wishes to thank Stefan Suhr for his clarifying comments on Theorem \ref{Suhrowe}.

\bibliographystyle{plain}
\bibliography{causality}{}

\end{document}